\documentclass[%
 reprint,
showpacs,
 amsmath,
 amssymb,
 aps,
floatfix,
]{revtex4-1}

\usepackage{graphicx}
\usepackage{dcolumn}
\usepackage{bm}

\usepackage{color}

\usepackage{ulem} 

\newcommand{\average}[1]{\left\langle#1\right\rangle}

\def\<{\langle}
\def\>{\rangle}
\def\({\left(}
\def\){\right)}
\def\[{\left[}
\def\]{\right]}

\def\e{\mathrm{e}}
\def\i{\mathrm{i}}

\def\der{\partial}

\def\Tr {{\mathrm{Tr}}}
\def\+{\dagger}
\def\rr{\bm{r}}
\def\R{\bm{R}}
\def\k{\bm{k}}

\begin{document}


\title{Orbital diamagnetic susceptibility in excitonic condensation phase}

\author{Koudai Sugimoto$^1$}
\author{Yukinori Ohta$^2$}%
\affiliation{%
$^1$Center for Frontier Science, Chiba University, Chiba 263-8522, Japan\\ 
$^2$Department of Physics, Chiba University, Chiba 263-8522, Japan
}%

\date{\today}

\begin{abstract}
We study the orbital diamagnetic susceptibility in excitonic condensation phase using the mean-field approximation for a two-band model defined on a square lattice.
We find that, in semiconductors, the excitonic condensation acquires a finite diamagnetic susceptibility due to spontaneous hybridization between the valence and the conduction bands, whereas in semimetals, the diamagnetic susceptibility in the normal phase is suppressed by the excitonic condensation.
We also study the orbital diamagnetic and Pauli paramagnetic susceptibilities of Ta$_2$NiSe$_5$ using a two-dimensional three-band model and find that the calculated temperature dependence of the magnetic susceptibility is in qualitative agreement with experiment.
\end{abstract}

\pacs{
71.30.+h, 
71.10.Fd, 
72.80.Ga 
71.35.Lk, 
}

\maketitle



\section{Introduction}

The spontaneous pair condensation of electrons and holes (excitons) in semiconductors or semimetals was predicted to occur as an exotic ground state of matter more than half a century ago~\cite{Mott1961PM, Knox1963SSP, Keldysh1965SPSS, Cloizeaux1965JPCS, Jerome1967PR, Halperin1968RMP}.
This phase is referred to as the excitonic (condensation) phase (EP).
Actual materials in the EP still are, however, being searched for because the exciton is charge neutral and, unlike in superconductivity, to detect the pair condensation experimentally is not straightforward.
One of the characteristic changes in the electronic structure at the EP transition is the band gap opening in semimetals and the band edge flattening in semiconductors, which angle-resolved photoemission spectroscopy (ARPES) experiments can detect.
It was thereby suggested that some materials such as Ta$_2$NiSe$_5$~\cite{Wakisaka2009PRL,Wakisaka2012JSNM,Kaneko2013PRB,Seki2014PRB} and $1T$-TiSe$_2$~\cite{Pillo200PRB, Kidd2002PRL, Cercellier2007PRL} are actually in the EP.  
In particular, for Ta$_2$NiSe$_5$, characteristic behaviors of the elastic constant, specific heat, ultrasonic attenuation rate, and NMR relaxation rate~\cite{Sugimoto2016PRB}, as well as 
the ARPES spectrum~\cite{,Kaneko2013PRB,Seki2014PRB}, were discussed in this respect.
A possible occurrence of a Fulde-Ferrell-Larkin-Ovchinnikov-type excitonic state in Ta$_2$NiSe$_5$ under high pressures was also discussed~\cite{Yamada2016JPSJ,Domon2016JPSJ}.  

Besides these physical quantities, it is known that a strong enhancement of the diamagnetic susceptibility below the excitonic transition temperature is observed in both Ta$_2$NiSe$_5$~\cite{DiSalvo1986JLCM} and $1T$-TiSe$_2$~\cite{DiSalvo1976PRB}, which suggests that  a fundamental relationship may exist between the excitonic condensation 
and diamagnetism.
In this paper, we therefore calculate the orbital diamagnetic susceptibility in the excitonic condensation phase and consider its physical significance, which we hope will shed some light on the excitonic condensation in real materials.

Orbital diamagnetic susceptibility in a periodic potential was first formulated by Peierls~\cite{Peierls1933ZP}, which was however applicable only to a single-band system.
Then, after much effort was made to extend the formula to multi-band systems, Fukuyama~\cite{Fukuyama1971PTP} succeeded to generalize the formula, writing it in a mathematically compact form.  This formula is applicable to tight-binding lattice models as well~\cite{Koshino2007PRB, Gomez2011PRL, Raoux2015PRB}, which therefore we will use in the present calculations.
It was recently pointed out \cite{Ogata2015JPSJ,Ogata2016JPSJ} that the use of Bloch wave functions, rather than the pure tight-binding lattice model, is important, the significance of which however we will leave for future study.
Because it is known that the effects of spin fluctuations hardly affect the diamagnetic susceptibility~\cite{Fukuyama1974PRB}, we expect that the formula will also be useful for EP.

In this paper, we will first introduce a two-orbital square-lattice model with an interorbital Coulomb interaction, which is a minimum model for the excitonic condensation with active spin degrees of freedom.
We will then obtain the EP of this model in the mean-field approximation and calculate the orbital diamagnetic susceptibility of this phase.
We will thereby show that, in semiconductors, the excitonic condensation acquires a finite diamagnetic susceptibility due to spontaneous hybridization between the valence and the conduction bands, whereas in semimetals, the diamagnetic susceptibility in the normal phase (NP) is suppressed by the excitonic condensation via the band gap opening.
We will clarify the origin of these behaviors by a simple model calculation.

We will also introduce a two-dimensional three-band model for describing the band structure near the Fermi level of Ta$_2$NiSe$_5$ and calculate the orbital diamagnetic and Pauli paramagnetic susceptibilities of this system, assuming the spin-singlet excitonic condensation.
We will show that the temperature dependence of the calculated total magnetic susceptibility is in qualitative agreement with experiment.

The rest of this paper is organized as follows.
In Sec.~\ref{sec:square_lattice_model}, we present our study on the two-orbital square-lattice model, where we obtain the mean-field solution for the EP of the model, calculate the orbital susceptibility, and discuss its significance to the excitonic condensation.
In Sec.~\ref{sec:tanise}, we present the two-dimensional tight-binding model of Ta$_2$NiSe$_5$ and, assuming the excitonic condensation, we calculate the orbital diamagnetic and Pauli paramagnetic susceptibilities of this system.
A summary of the paper is given in Sec.~\ref{sec:summary}.


\section{Study on the square-lattice model}\label{sec:square_lattice_model}

\subsection{Excitonic condensation}\label{sec:ex_cond}

\begin{figure}
\includegraphics[width = \linewidth]{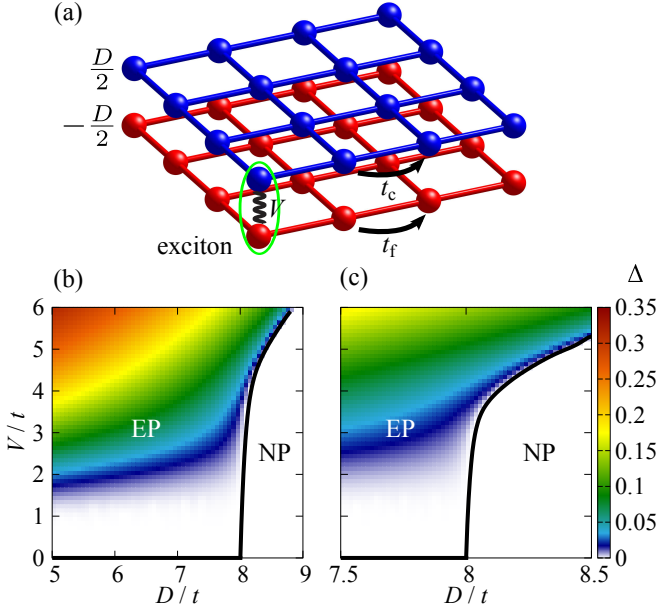}
\caption{(Color online) 
(a) Schematic of the two-orbital model defined on the two-dimensional square lattice.
(b) Phase diagram of the EP on the $(D/t,V/t)$ plane with $t_{\rm f}=-t_{\rm c}=t$ where the contour plot of the order parameter $\Delta$ is shown.
The phase boundary between the EP and the NP is indicated by a thick solid line.
(c) The same as in (b), but the region near the phase boundary is enlarged.
}\label{fig:op_phase}
\end{figure}

Let us first introduce a two-orbital model defined on the two-dimensional square lattice [see Fig.~\ref{fig:op_phase}(a)], where the $f$ and $c$ orbitals form the valence and conduction bands with hopping integrals $t_{\rm f}$ and $t_{\rm c}$, respectively, which are separated by the energy level splitting $D$.
There is no hopping of electrons between the $f$ and the $c$ orbitals, but an interorbital repulsive interaction $V$ acts between two electrons in the $f$ and $c$ orbitals.
This is a minimum lattice model for the excitonic condensation.
The Hamiltonian is written as $H = H_0 + H_V$ with
\begin{align}
&H_0 = \sum_{\< i, j \>, \sigma} \( t_{\rm f} f_{i, \sigma}^\+ f_{j, \sigma} + t_{\rm c} c_{i, \sigma}^\+ c_{j, \sigma} \)
\label{eq:H_0}
\notag \\
&~~~~~~~~~~~~~~+ \frac{D}{2} \sum_{i, \sigma} \( c_{i, \sigma}^\+ c_{i, \sigma} - f_{i, \sigma}^\+ f_{i, \sigma} \),
\\
&H_V = V \sum_{i, \sigma, \sigma'} f_{i, \sigma}^\+ f_{i, \sigma} c_{i, \sigma'}^\+ c_{i, \sigma'}, 
\label{eq:H_V}
\end{align}
where $f_{i, \sigma}$ ($f_{j, \sigma}^\+$) is the annihilation (creation) operator of an electron with spin $\sigma$ in the $f$ orbital at site $i$ and $c_{i, \sigma}$ ($c_{j, \sigma}^\+$) is that in the $c$ orbital.
The symbol $\<i,j\>$ stands for the nearest-neighbor pair of sites $i$ and $j$.

Defining the order parameter of the spin-singlet EP as
\begin{equation}
 \Delta = \average{c^{\+}_{i, \sigma} f_{i, \sigma'}} \delta_{\sigma, \sigma'}, 
\end{equation}
we rewrite Eq.~(\ref{eq:H_V}) into 
\begin{equation}
 H_V^{\rm MF}
	= - V \( \Delta  \sum_{i, \sigma} f_{i, \sigma}^\+ c_{i, \sigma}
		+ {\rm H.c.} \),
\label{eq:zHbORi6s}
\end{equation}
in the mean-field approximation.  Here, we neglect the intraorbital terms containing $\average{c_{i, \sigma}^\+ c_{i, \sigma}}$ or $\average{f_{i, \sigma}^\+ f_{i, \sigma}}$ because we do not consider the other ordered phases such as spin-density-wave and charge-density-wave phases in the present study.
Introducing the Fourier transformations
$f_{i, \sigma} = \frac{1}{\sqrt{N}}\sum_{\k} \e^{\i \k \cdot \rr_{i}} f_{\k, \sigma}$ and
$c_{i, \sigma} = \frac{1}{\sqrt{N}}\sum_{\k} \e^{\i \k \cdot \rr_{i}} c_{\k, \sigma}$,
where $N$ is the number of the unit cells, we obtain the mean-field Hamiltonian,
\begin{equation}
 H^{\rm MF}
	= \sum_{\k, \sigma}
	\Psi_{\k, \sigma}^{\+}
	{\cal H}_{\k}
	\Psi_{\k, \sigma},
\end{equation}
in the spinor representation 
$\Psi_{\k, \sigma}^{\+} = \( c_{\k, \sigma}^{\+}, f_{\k, \sigma}^{\+} \)$ with 
\begin{align}
 &{\cal H}_{\k}=
\notag \\
	&\begin{pmatrix}
		2 t_{\rm c} \( \cos {k_x a} + \cos {k_y a} \)+\frac{D}{2} & -V \Delta^* \\
		-V \Delta & 2 t_{\rm f} \( \cos {k_x a} + \cos {k_y a} \)-\frac{D}{2}
	\end{pmatrix}, 
\end{align}
where $a$ is the lattice constant.
Note that the Hartree shift is excluded since we neglect the intraorbital terms.

\subsection{Orbital susceptibility}\label{sec:orb_sus}

Applying a uniform magnetic field perpendicular to the lattice plane, the orbital susceptibility of our system is given by~\cite{Gomez2011PRL} 
\begin{multline}
 \chi_{\rm orb}
 	= \frac{\mu_{\rm B}^2}{2 a_{\rm B}^4 Ry^2 \beta N} \(\frac{\hbar}{e}\)^4
		\sum_{\k, m}
		\Tr \Bigl(
			2 {\cal G} \, j^{x} \, {\cal G} \, {j}^{y}
				{\cal G} \, {j}^{x} \, {\cal G} \, {j}^{y}
\\
			- {\cal G} \, {\tau}_{x, y} \, {\cal G} \, {j}^{y} \, {\cal G} \, {j}^{x}
			- {\cal G} \, {\tau}_{x, y} \, {\cal G} \, {j}^{x} \, {\cal G} \, {j}^{y} \Bigr),
\label{eq:cO9N6lcF}
\end{multline}
or equivalently by \cite{Raoux2015PRB}
\begin{multline}
 \chi_{\rm orb}
 	= \frac{\mu_{\rm B}^2}{6 a_{\rm B}^4 Ry^2 \beta N} \(\frac{\hbar}{e}\)^4
		\sum_{\k, m}
		\Tr \Bigl[
			{\cal G} \, {\tau}_{x, x} \, {\cal G} \, {\tau}_{y, y} +
\\
			 {\cal G} \, {\tau}_{x, y} \, {\cal G} \, {\tau}_{x,y}
			- 4 \( {\cal G} \, j^{x} \, {\cal G} \, {j}^{y} \,
				{\cal G} \, {j}^{x} \, {\cal G} \, {j}^{y}
				- {\cal G} \, j^{x} \, {\cal G} \, {j}^{y} \,
				{\cal G} \, {j}^{x} \, {\cal G} \, {j}^{y} \) \Bigr],
\end{multline}
where $\mu_{\rm B} = e \hbar / 2mc$ is the Bohr magneton, $a_{\rm B} = \hbar^2 / me^2$ is the Bohr radius, and $Ry = e^2 / 2 a_{\rm B}$ is the Rydberg constant.
Here, we define the electric current,
\begin{equation}
\bm{j}= - \frac{e}{\hbar} \frac{\der {\cal H}_{\k}}{\der \k},
\end{equation}
and stress tensor,
\begin{equation}
\tau_{\alpha, \beta} = - \( \frac{e}{\hbar} \)^2 \frac{\der^2 {\cal H}_{\k}}{\der k_{\alpha} \der k_{\beta}}. 
\end{equation}
${\cal G}$ is the temperature Green's function written as
\begin{equation}
 {\cal G} (\k, \i \omega_m)
	= \Big[ \i \hbar \omega_m - \( {\cal H}_{\k} - \mu \) \Big]^{-1},
\end{equation}
where $\omega_m = \( 2m + 1 \) \pi  / \beta \hbar$ is the Matsubara frequency with reciprocal temperature $\beta = 1 / k_{\rm B}T$ and $\mu$ is the chemical potential.
$k_{\rm B}$ is the Boltzmann constant.

For the $\k$-summation, we divide the Brillouin zone into $400 \times 400$ meshes in the mean-field self-consistent calculations and $1000 \times 1000$ meshes in the susceptibility calculations.
The summation over the Matsubara frequencies is carried out with the usual analytical continuation technique.

\subsection{Results for the square-lattice model}\label{sec:result}

We assume a particle-hole symmetric situation $t = t_{\rm f} = - t_{\rm c}$ and a number of electrons at half filling (two electrons per site), so that we can set the chemical potential to be zero.
We thus have a direct-gap semiconductor for $D/t>8$ and a semimetal for $D/t<8$ at $V/t=0$.
Figure \ref{fig:op_phase}(b) shows the calculated phase diagram of our model in the mean-field approximation.
This phase diagram is enlarged near the semimetal-semiconductor phase boundary in Fig.~\ref{fig:op_phase}(c).
We find that, in the semimetallic region $D/t<8$, the EP persists down to an infinitesimal value of $V/t$, whereas in the semiconducting region $D/t > 8$, it vanishes at a finite value of $V/t$.
Thus, a comparatively large value of the interorbital Coulomb interaction is required for the excitonic condensation in the semiconducting region.
This result is in apparent contrast to the result of the electron gas model~\cite{Bronold2006PRB, Hulsen2006PhysicaB}, where the EP survives well above the semimetal-semiconductor transition point.
This contrast may be understood because we assume a constant value of the Coulomb interaction $V$ in the lattice model, whereas in the gas model, the interaction is screened in the semimetallic region but not in the semiconducting region.

We then calculate the orbital susceptibility as \cite{Raoux2015PRB}
\begin{align}
 \chi_{\rm orb}
	&= \frac{2\mu_{\rm B}^2 a^4 t^2}{3 a_{\rm B}^4 Ry^2 N} \sum_{\k} \cos k_x a \cos k_y a 
\notag \\
		&~~~~~\times \Biggl\{ \( 1 - \frac{V^2\left|\Delta\right|^2}{\epsilon_k^2}\)
			 \( n'_{\rm F} (\epsilon_k) + n'_{\rm F} (-\epsilon_k) \)
\notag \\
	&~~~~~~~~~~~~+  \frac{V^2 \left| \Delta\right| ^2}{\epsilon_k^2} \( \frac{n_{\rm F} (\epsilon_k) - n_{\rm F} (-\epsilon_k)}{\epsilon_k}\)
		\Biggr\},
\end{align}
with
\begin{equation}
 \epsilon_{\k}
	= \sqrt{\[ -2t\(\cos k_x a + \cos k_y a \) + \frac{D}{2} \]^2 + V^2 \left| \Delta \right|^2}, 
\end{equation}
where $n_{\rm F} (\epsilon) $ is the Fermi distribution function and $n_{\rm F}' (\epsilon)$ is its derivative.

\begin{figure}
\includegraphics[width = \linewidth]{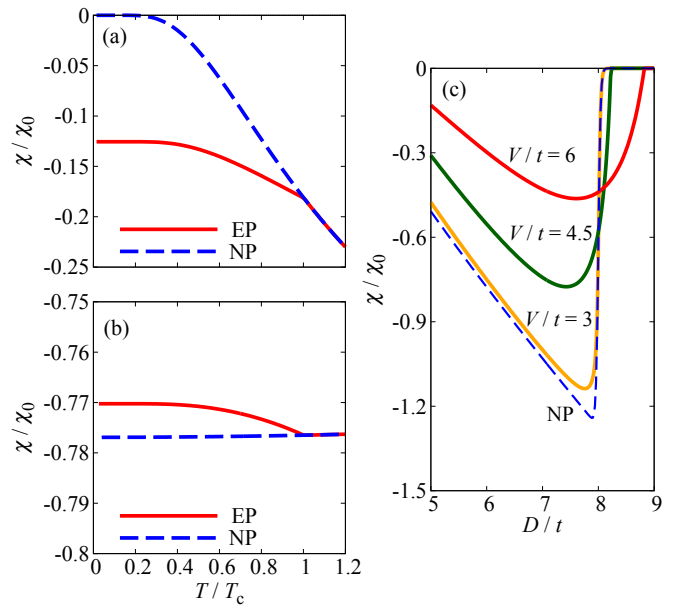}
\caption{(Color online) 
Calculated diamagnetic susceptibility
(a) as a function of temperature at $D/t = 8.2$ and $V/t=4.5$ (semiconducting region),
(b) as a function of temperature at $D/t = 6$ and $V/t=2.5$ (semimetallic region), and
(c) as a function of the level splitting $D/t$ at low temperature $k_{\rm B} T / t = 0.01$.
The dashed lines are for the NP, and the solid lines are for the EP.
}\label{fig:diamag_ep}
\end{figure}

The calculated results for the temperature dependence of the orbital susceptibility are shown in Fig.~\ref{fig:diamag_ep} where we define a constant $\chi_0 = \frac{\mu_{\rm B}^2 a^4 t}{6 a_{\rm B}^4 Ry^2 N}$.
Let us first discuss the semiconducting case [see Fig.~\ref{fig:diamag_ep}(a)] where we assume the parameter values $D/t=8.2$ and $V/t=4.5$, so that we obtain the transition temperature $k_{\rm B} T_{\rm c}/t = 0.056$.
Above the transition temperature, the orbital susceptibility is negative, indicating that the system is diamagnetic.
As the temperature decreases, the EP transition occurs, at which the orbital susceptibility shows a kink.
Decreasing the temperature further, we find that the diamagnetic susceptibility is much enhanced in the EP, compared with the NP.
At zero temperature, the orbital susceptibility remains negative (diamagnetic), whereas it vanishes in the NP.
Next, let us discuss the semimetallic case [see Fig.~\ref{fig:diamag_ep}(b)] where we assume $D/t=6$ and $V/t=2.5$, so that we obtain $k_{\rm B}T_{\rm c}/t=0.065$.
In the NP, the orbital susceptibility is largely negative (strongly diamagnetic) and almost temperature independent.
As the temperature decreases, the orbital susceptibility shows a kink at the excitonic transition, and below the transition temperature, the diamagnetism is slightly weakened in the EP, compared with the NP.

The orbital susceptibility calculated as a function of $D/t$ at low-temperature $k_{\rm B} T / t = 0.01$ is shown in Fig.~\ref{fig:diamag_ep}(c) where we find that the essential difference occurs between the semiconducting and the semimetallic regions.
In the semiconducting region $D/t>8$, the orbital susceptibility is almost zero in the NP, and only for large values of $V/t$ where the system goes into the EP, does the diamagnetism appear.
In the semimetallic region $D/t<8$, on the other hand, the orbital susceptibility is largely negative (strongly diamagnetic) already in the NP, and the diamagnetism is weakened when the system goes into the EP with increasing $V/t$.

We note here that exactly the same results for the orbital susceptibility as above are obtained when we assume the spin-triplet excitonic condensation.
Also noted is that no change occurs in our results for the orbital susceptibility in the indirect gap situation $t_{\rm f}=t_{\rm c}$.

\begin{figure}
\includegraphics[width = \linewidth]{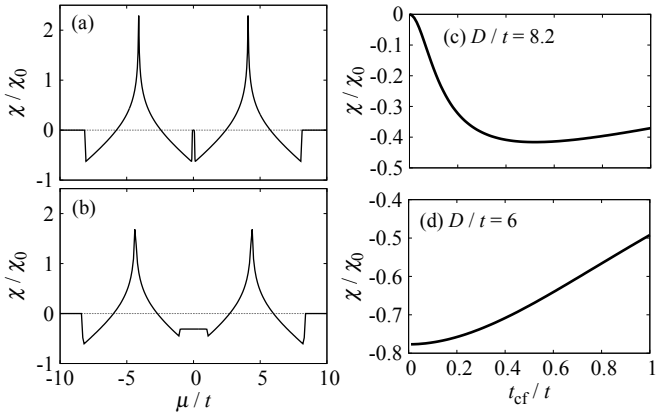}
\caption{
Left panels: calculated orbital susceptibility as a function of the chemical potential $\mu/t$ in the semiconducting NP ($D/t=8.2$) at low-temperature $k_{\rm B} T/t = 0.01$
(a) without the $c$-$f$ hybridization $t_{\rm cf}=0$ and (b) with the $c$-$f$ hybridization $t_{\rm cf}/t=1$.  
Right panels: calculated orbital susceptibility as a function of the $c$-$f$ hybridization $t_{\rm cf}/t$ at $k_{\rm B} T/t = 0.01$,
(c) in the semiconducting NP ($D/t=8.2$) and
(d) in the semimetallic NP ($D/t=6$).
No excitonic condensation is taken into account.
}\label{fig:diamag_tcf}
\end{figure}

Now, let us clarify the origin of the above-discussed behaviors of the orbital susceptibility.
To this end, we introduce a hybridization $t_{\rm cf}$  between the $c$ and the $f$ orbitals artificially, without taking into account the excitonic condensation; i.e., we add a $c$-$f$ hybridization term,
\begin{equation}
 H_{\rm cf} = t_{\rm cf} \sum_{i, \sigma} \( c_{i, \sigma}^\+ f_{i, \sigma} + f_{i, \sigma}^\+ c_{i, \sigma} \)
 \label{eq:o5MCrUzx}
\end{equation}
to the noninteracting Hamiltonian [Eq.~(\ref{eq:H_0})] but we neglect the interaction term [Eq.~(\ref{eq:H_V})].
The model remains electron-hole symmetric.
We thus calculate the orbital susceptibility using the formula given in Sec.~\ref{sec:orb_sus}.

First, we consider the semiconducting case ($D/t=8.2$) in the absence of the $c$-$f$ hybridization $t_{\rm cf}/t=0$.
The orbital susceptibility calculated as a function of the chemical potential $\mu/t$ is shown in Fig.~\ref{fig:diamag_tcf}(a).
Here, the $c$ and $f$ orbitals are completely independent so that it is clear that the orbital susceptibility vanishes when $\mu/t$ is in the semiconducting gap, i.e., the electrons in the filled $f$ band cannot move with an infinitesimal magnetic field.
The peaks at $\mu/t=\pm 4.1$ are due to the van Hove singularity of the present model.
Now, introducing a finite value of $t_{\rm cf}$, we find that the system acquires the diamagnetic susceptibility even at $\mu/t=0$ [see Fig.~\ref{fig:diamag_tcf}(b)], where the band gap remains open.
This result may be understood because the electrons in the filled valence band become mobile via the electron hopping $t_{\rm cf}$ (or $c$-$f$ hybridization) under an infinitesimal magnetic field.

In Fig.~\ref{fig:diamag_tcf}(c), we show the $t_{\rm cf}$ dependence of the orbital susceptibility in the semiconducting case.
We find that the system acquires the diamagnetism with increasing $t_{\rm cf}$ as discussed above.
However, the diamagnetism is weakened again for very large values of $t_{\rm cf}$ because the band gap in this situation becomes too large for the electrons to move easily.
In the semimetallic case, of which the result is shown in Fig.~\ref{fig:diamag_tcf}(d), we find that the orbital susceptibility, which is largely negative even at $t_{\rm cf} = 0$ as discussed above, is suppressed with increasing $t_{\rm cf}$.  This is because the band gap opens at $t_{\rm cf}>0$ and the gap size increases with increasing $t_{\rm cf}$, so that the electrons become less mobile.

The same discussion as above can be applied to the interpretation of the behavior of the orbital susceptibility calculated in the EP because the essential feature of the excitonic 
condensation is the spontaneous hybridization between the valence and the conduction bands.
Namely, the excitonic order parameter in Eq.~(\ref{eq:zHbORi6s}) plays exactly the same role as $t_{\rm cf}$ in Eq.~(\ref{eq:o5MCrUzx}).

\begin{figure}
\includegraphics[width=0.7\linewidth]{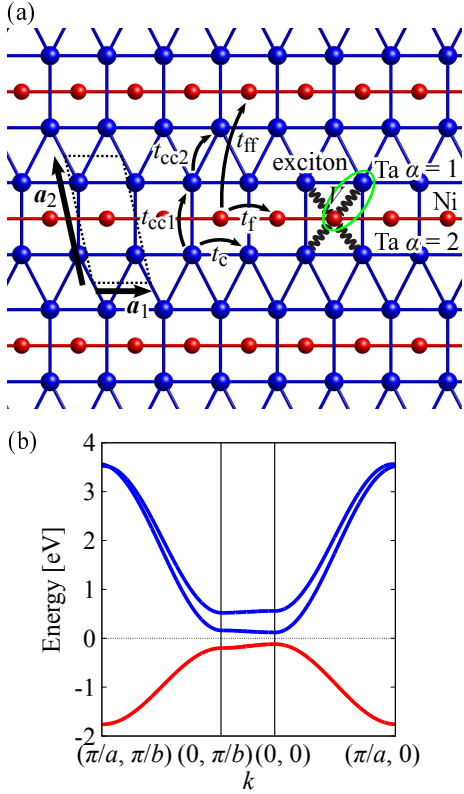}
\caption{(Color online)
(a) Schematic of the two-dimensional three-band model of Ta$_2$NiSe$_5$, where the dotted parallelogram stands for the unit cell.
(b) The tight-binding band structure of the model shown in (a).
The valence band comes from the Ni ions, and the two conduction bands come from the Ta ions.
$a$ and $b$ are the lattice constants.
}\label{fig:tns2d}
\end{figure}

\section{Magnetic susceptibility of Ta$_2$NiSe$_5$}\label{sec:tanise}

Let us apply our theory to Ta$_2$NiSe$_5$, which is a candidate material for the spin-singlet excitonic condensation.
Because the orbital susceptibility calculation requires the system of more than one dimension, we extend the one-dimensional three-chain model proposed~\cite{Kaneko2013PRB, Sugimoto2016PRB} to a two-dimensional one where the interchain hopping parameters are introduced as shown in Fig.~\ref{fig:tns2d}(a).
The noninteracting tight-binding Hamiltonian reads 
\begin{align}
&H_0=t_{\rm c} \sum_{j, \alpha, \sigma} \sum_{\bm{\delta} = \pm \bm{a}_1}
			c_{\R_j + \bm{\delta}, \alpha, \sigma}^\+ c_{\R_j, \alpha, \sigma}
+ \varepsilon_{\rm c} \sum_{j, \alpha, \sigma} c_{j, \alpha, \sigma}^\+ c_{j, \alpha, \sigma}
\notag \\
		&+ t_{\rm f} \sum_{j, \sigma} \sum_{\bm{\delta} = \pm \bm{a}_1}
			f_{\R_j + \bm{\delta}, \sigma}^\+ f_{j, \sigma}
		+ \varepsilon_{\rm f} \sum_{j, \sigma} f_{j, \sigma}^\+ f_{j, \sigma}
\notag \\
		&+ t_{\rm cc1} \sum_{j, \sigma}
			\( c_{\R_j - \bm{a}_1, 2, \sigma}^\+ c_{j, 1, \sigma} + {\rm H.c.}\)
\notag \\
		&+ t_{\rm cc2} \sum_{j, \alpha, \sigma}
			\( c_{\R_j + \bm{a}_2, 2, \sigma}^\+ c_{j, 1, \sigma}
			+ c_{\R_j + \bm{a}_2 - \bm{a}_1, 2, \sigma}^\+ c_{j, 1, \sigma}
			+{\rm H.c.} \)
\notag \\
		&+ t_{\rm ff} \sum_{j, \alpha, \sigma}
			\( f_{\R_j + \bm{a}_2, \sigma}^\+ f_{j, \sigma}
			+ f_{\R_j + \bm{a}_2 + \bm{a}_1, \sigma}^\+ f_{j, \sigma} 
			+{\rm H.c.} \), 
\end{align}
where $t_{\rm c}$ is the hopping integral along the Ta chains, $t_{\rm f}$ is that along the Ni chains,
$t_{\rm cc1}$ and $t_{\rm cc2}$ are the interchain hopping integrals between the Ta chains, and
$t_{\rm ff}$ is the interchain hopping integral between the Ni chains.  These are illustrated in Fig.~\ref{fig:tns2d}(a).
The annihilation operator of an electron with spin $\sigma$ in the $\alpha$th Ta chain is defined as 
$c_{j, \alpha, \sigma}$ or $c_{\R_{j}, \alpha, \sigma}$ and that in the Ni chain is defined as 
$f_{j, \sigma}$ or $f_{\R_{j}, \sigma}$, where $\R_{j}$ is the position of the $j$th unit cell.  
From the band structure calculation~\cite{Kaneko2013PRB}, we set 
$t_{\rm c} = -0.8$,
$t_{\rm f} = 0.4$,
$t_{\rm cc1} = -0.02$,
$t_{\rm cc2} = -0.1$,
$t_{\rm ff} = 0.01$, and
$\varepsilon_{\rm c} - \varepsilon_{\rm f} = 2.95$ in units of eV.
The band dispersions of this model are shown in Fig.~\ref{fig:tns2d}(b).  
The primitive cell vectors are given by $\bm{a}_1 = (a, 0)$ and $\bm{a}_2 = (-a/2, b)$, where $a = 3.496$ and $b = 7.820$~\AA~are estimated from experiment~\cite{Sunshine1985IC}.
Note that the doubly degenerate conduction bands in the three-chain model~\cite{Kaneko2013PRB} split into two due to the interchain hopping integral between the Ta chains.  

We also include the intersite repulsion term between Ni and Ta ions, just as in the three-chain 
model~\cite{Kaneko2013PRB,Sugimoto2016PRB}, which is defined as 
\begin{align}
&H_{V} = V \sum_{j, \sigma, \sigma'} n_{j, \sigma}^{\rm f} \times
\notag \\
&~~~~\( n_{j, 1, \sigma'}^{\rm c} + n_{\R_j + \bm{a}_1, 1, \sigma'}^{\rm c} 
+ n_{j, 2, \sigma'}^{\rm c} + n_{\R_{j} - \bm{a}_{1}, 2, \sigma'}^{\rm c} \), 
\label{eq:H_V_tns}
\end{align}
where $n_{j, \alpha, \sigma}^{\rm c} = c_{j, \alpha, \sigma}^\+ c_{j, \alpha, \sigma}$ and 
$ n_{j, \alpha, \sigma}^{\rm f} = f_{j, \sigma}^\+ f_{j, \sigma}$.  $V$ is the strength of this interaction.
The intrasite repulsion terms in the Ni and Ta ions are neglected because of the same reasons given in Sec.~\ref{sec:ex_cond}.
The electron-phonon coupling term is required to explain the lattice distortion that occurs at the EP transition in Ta$_2$NiSe$_5$.
However, because the mean-field Hamiltonian of the system is written in terms of a sum of the order parameter of the lattice distortion and the order parameter of the excitonic condensation as is evident in Ref.~\onlinecite{Kaneko2013PRB}, the electron-phonon coupling term contributes to the orbital susceptibility just as the $V$ term does.
This means that the two contributions to the orbital susceptibility cannot be distinguished within the framework of the theory so that hereafter we only refer to the $V$ term as a representative of both the $V$ term and the electron-phonon coupling term.

Then, defining the excitonic order parameter as
\begin{align}
 \Delta = \average{c^{\+}_{i, \alpha, \sigma} f_{i, \sigma'}} \delta_{\sigma, \sigma'} 
&= \average{c^{\+}_{\bm{R}_i+\bm{a}_1, 1, \sigma} f_{i, \sigma'}} \delta_{\sigma, \sigma'}
\notag \\
&= \average{c^{\+}_{\bm{R}_i-\bm{a}_1, 2, \sigma} f_{i, \sigma'}} \delta_{\sigma, \sigma'}, 
\end{align}
we apply the mean-field approximation to Eq.~(\ref{eq:H_V_tns}), just as in Eq.~(\ref{eq:zHbORi6s}), 
and solve the gap equation self-consistently.
We thus find a transition temperature of $605$~K at $V = 0.9$~eV, which is considerably larger than the experimental value of 328 K \cite{DiSalvo1986JLCM}; the discrepancy may be attributed to the mean-field approximation ignoring quantum fluctuations.
We then apply the formula discussed in Sec.~\ref{sec:orb_sus} and calculate the orbital susceptibility of this phase.

The Pauli paramagnetic susceptibility, which may be affected by the excitonic condensation and 
can have a strong temperature dependence, may also contribute to the temperature dependence 
of the magnetic susceptibility of Ta$_2$NiSe$_5$.
We therefore calculate the spin susceptibility as well, which is given by
\begin{equation}
 \chi_{\rm spin}
	= -\frac{2\mu_{\rm B}^2}N \sum_{\k, \epsilon}
	\frac{\beta}{4 \cosh^2 \big( \beta E_{\k, \epsilon, \sigma - \mu} / 2 \big)},
\end{equation}
where $E_{\k, \epsilon, \sigma}$ is the eigenenergy of the gap equation with the wave-vector 
$\k$, band $\epsilon$, and spin $\sigma$~\cite{Mahan2000}.  

\begin{figure}
\includegraphics[width=0.7\linewidth]{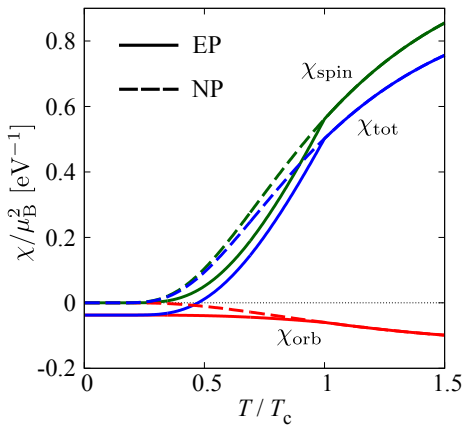}
\caption{(Color online) 
Calculated temperature dependence of the magnetic susceptibility of our model in the EP, solid lines where the orbital $\chi_{\rm orb}$ and spin $\chi_{\rm spin}$ contributions 
are separately shown, together with the total magnetic susceptibility $\chi_{\rm tot}$.
The susceptibilities in the NP are also shown by the dashed lines.
}\label{fig:tns_suscept}
\end{figure}

Figure~\ref{fig:tns_suscept} shows the calculated temperature dependence of the orbital susceptibility $\chi_{\rm orb}$, spin susceptibility $\chi_{\rm spin}$, and total susceptibility $\chi_{\rm tot} = \chi_{\rm orb} + \chi_{\rm spin}$ in the EP, where we use the two-dimensional three-band model discussed above.
We find that the orbital susceptibility shows diamagnetism and has a typical temperature dependence in the semiconducting phase as discussed in Sec.~\ref{sec:result} because this system is a direct-gap semiconductor.
However, the contribution of the orbital susceptibility is rather small.
This is because the system is quasione dimensional so that the component coming from the electric current perpendicular to the chains is much smaller than the component coming from the electric current parallel to the chains.
The spin susceptibility, on the other hand, has large values at high temperatures and decreases rapidly below the excitonic transition temperature.

We do not include the large diamagnetic contributions from the core electrons as well as the Van Vleck susceptibility, which are important in the total magnetic susceptibility~\cite{Fazekas1999}.
However, these contributions are almost temperature independent and lead to a uniform negative shift to the magnetic susceptibility in total.
Our result indicates that the spin susceptibility and orbital susceptibility cooperatively enhance the diamagnetism in the EP of Ta$_2$NiSe$_5$, which is qualitatively consistent with experiment~\cite{DiSalvo1986JLCM}.
Effects of electron correlations as well as a recent development in the theory of diamagnetism~\cite{Ogata2015JPSJ,Ogata2016JPSJ}, which are neglected in the present calculations, should be taken into account for more quantitative discussions, but we believe that the essential features of the temperature dependence of the magnetic susceptibility of Ta$_2$NiSe$_5$ assuming the excitonic condensation are obtained in the present calculations.


\section{Summary}\label{sec:summary}

We have studied the orbital diamagnetic susceptibility in the excitonic condensation phase using the mean-field approximation for the interacting tight-binding lattice models. 
We calculated the orbital susceptibility for the two-band model defined on the square lattice, and found that, in semiconductors, the excitonic condensation acquires a finite diamagnetic susceptibility, whereas in semimetals, the diamagnetic susceptibility in the NP is suppressed by the excitonic condensation.
We showed that these results can be interpreted in terms of the hybridization between the valence and the conduction bands; in semiconductors, the electrons in the valence band become mobile via the spontaneous hybridization with the conduction band so that the system acquires the diamagnetism when the excitonic condensation occurs, whereas in semimetals, the system is diamagnetic in the NP and the spontaneous hybridization between the valence and the conduction bands leads to the band gap opening, which suppresses the diamagnetism.

We also studied the orbital diamagnetic and Pauli paramagnetic susceptibilities of Ta$_2$NiSe$_5$ using the two-dimensional three-band model and found that the spin and orbital susceptibilities cooperatively lead to the rapid decrease in the magnetic susceptibility due to the excitonic condensation, which is in qualitative agreement with experiment.

\begin{acknowledgments}
We thank Professor H. Fukuyama for directing our attention to the present issue.  
We also thank M. Itoh, Y. Kobayashi, T. Mizokawa, and H. Takagi for discussing experimental aspects of Ta$_2$NiSe$_5$ and T. Kaneko and M. Ogata for theoretical aspects.  
This work was supported in part by Grants-in-Aid for Scientific Research from JSPS (Grants No.~26400349 and No. 15H06093) of Japan.
\end{acknowledgments}

\nocite{*}



\end{document}